\documentclass[11pt]{article}
\usepackage[margin=1in]{geometry}
\usepackage{amsmath,amssymb,amsthm}
\usepackage{mathtools}
\usepackage{algorithm}
\usepackage{algpseudocode}
\usepackage{tikz}
\usepackage{natbib}
\usepackage{microtype}
\usepackage[hidelinks]{hyperref}
\allowdisplaybreaks
\newtheorem{theorem}{Theorem}
\newtheorem{lemma}{Lemma}
\newtheorem{corollary}{Corollary}

\newcommand{\size}[1]{\ensuremath{|#1|}}
\def\A{\mathcal{A}}
\def\B{\mathcal{B}}
\def\F{\mathcal{F}}
\def\O{\mathcal{O}}
\def\P{\mathcal{P}}
\def\S{\mathcal{S}}
\def\OPT{\mathrm{opt}}

\title{A Local-to-Global Approximation Framework for Covering Disjointness-Compliable Set Families}
\author{Jingyang Zhao\thanks{Kyung Hee University. Email: \texttt{jingyangzhao1020@gmail.com}.}
\and
Mingyu Xiao\thanks{University of Electronic Science and Technology of China. Email: \texttt{myxiao@gmail.com}.}}
\date{}

\begin{document}
\maketitle
\begin{abstract}
Many network design problems can be formulated as covering a family of vertex sets by edges.
We develop a local-to-global approximation framework for disjointness-compliable families with polynomial-time membership tests.
A class of vertices is deficient if it belongs to the family and feasible otherwise. Our local problem seeks a minimum-cost tree that connects one deficient class to feasible classes so that their union is feasible.
If this local problem admits a polynomial-time $\alpha$-approximation, then the covering problem admits a polynomial-time $(1+\varepsilon)(\sqrt{2}+\sqrt{\alpha})^2$-approximation for every fixed $\varepsilon>0$.
Our algorithm first connects deficient vertices that are difficult to separate and then handles regions containing one deficient class at a time.
To bound the total local cost, we compare the local trees with a feasible forest. We delete edges of this forest after using them in the bound, so each is counted at most once even when regions overlap.
As applications, we obtain polynomial-time $(8+\varepsilon)$-approximation algorithms for generalized point-to-point connection and multiroot quota tree, including multiroot $k$-minimum spanning tree.
To the best of our knowledge, these are the first polynomial-time constant-factor approximations for both problems.
We also obtain a randomized approximation for multiroot covering Steiner that has the same asymptotic ratio as the known approximation for its single-root version.
\end{abstract}

\section{Introduction}\label{sec:introduction}
Network design problems arise when a planner must install connections to satisfy requirements at different locations.
For example, a communication network may need to connect prescribed pairs of terminals, while a distribution network may need to connect demand locations to sufficient supply.
The planner pays for the selected connections and seeks to minimize their total cost.
When an edge is paid for once, several connections or shipments can share it without additional installation cost~\citep{journal:even:TALG:05,journal:hajiaghayi:TOCS:16,preprint:nutov:arXiv:25}.
Solving each requirement separately may overlook these savings and lead to a more expensive network.

The requirements can take different forms.
Some specify which pairs of vertices must be connected; others require each connected component to contain sufficient supply or prize.
In either case, a set of vertices may need at least one selected edge leading to the rest of the network.
For instance, a group of locations whose demand exceeds its supply must connect to other locations to receive enough supply.
This view expresses the requirements as a family of vertex sets that must be covered by edges~\citep{journal:goemans:SICOMP:95,preprint:nutov:arXiv:25}.
We study a general approximation framework for such set-family covering problems.

\subsection{Set-Family Covering}\label{sec:intro-definitions}
Let $G=(V,E)$ be an undirected graph with nonnegative edge costs $c(e)$, and let $\F\subseteq2^V$ be a family of nonempty vertex sets.
An edge \emph{covers} a set $S$ if it has exactly one endpoint in $S$.
The goal is to find a minimum-cost edge set that covers every set in $\F$.
The family can be specified implicitly; we assume that membership in $\F$ can be checked in polynomial time.
For example, in the Steiner forest problem (SF), we must connect each prescribed pair of terminals.
We can express this requirement by placing in $\F$ every set that contains exactly one endpoint of at least one pair.
An edge set covers this family if and only if it connects every prescribed pair.

We focus on \emph{disjointness-compliable} families~\citep{journal:goemans:SICOMP:95,preprint:nutov:arXiv:25}.
For every $S\in\F$ and $B\subseteq S$, at least one of $B$ and $S\setminus B$ must belong to $\F$.
Equivalently, for any $A,B\subseteq V$, if $A,B\notin\F$ and $A\cap B=\emptyset$, then $A\cup B\notin\F$.
We refer to the problem of covering such a family as the disjointness-compliable family covering problem (DCFC).
An important special case arises when $\F$ is also symmetric: for every $S\subseteq V$, we have $S\in\F$ if and only if $V\setminus S\in\F$.
Such a family is called proper.
\citet{journal:goemans:SICOMP:95} developed a primal-dual $2$-approximation algorithm for covering proper families, including the family associated with the SF.
Our setting also allows families that are not symmetric, so the result for proper families does not directly apply.

To develop an algorithm for this broader setting, we use the disjointness-compliable property to describe a cover through its connected components.
We call a set in $\F$ deficient and a set outside $\F$ feasible.
An edge set covers $\F$ if and only if each of its components is feasible~\citep[see the proof of Theorem~2.3]{journal:goemans:SICOMP:95}.

\subsection{Applications and Related Work}\label{sec:related}
Many network design problems can be expressed as instances of the DCFC.
We first introduce several of these problems and their known approximation results.

\paragraph{Steiner Forest.}
The primal-dual algorithm of \citet{journal:goemans:SICOMP:95} achieves approximation ratio $2$ relative to the standard cut relaxation.
More recently, \citet{conference:gupta:STOC:26} obtained a $1.994$-approximation relative to the cost of an optimal forest.

\paragraph{Quota Tree.}
In the rooted quota-tree problem, each vertex has a nonnegative prize, and a tree containing a specified root must collect at least a given quota.
The rooted $k$-minimum spanning tree problem ($k$-MST) is the special case in which each vertex has unit prize and the quota is $k$.
Thus, the tree must contain at least $k$ vertices, including the root.
\citet{conference:garg:STOC:05} obtained a $2$-approximation for $k$-MST; see also \citet{conference:breen:SOSA:23}.
For the rooted quota-tree problem with nonnegative vertex prizes, \citet{conference:friggstad:WADS:25} established a $(2+\varepsilon)$-approximation for every fixed $\varepsilon>0$.

\paragraph{Point-to-Point Connection.}
The point-to-point connection problem (P2P) seeks a minimum-cost edge set connecting sources to destinations~\citep{journal:li:DAM:92}.
In the fixed-destination version, each source must be connected to its prescribed destination, yielding the SF.
In the nonfixed version, there are equally many sources and destinations, but their matching is not prescribed; each component must contain the same number of each.
In the generalized P2P (GP2P), each vertex $v$ has an integer charge $\phi(v)$: a positive charge represents supply and a negative charge represents demand.
For $S\subseteq V$, let $\phi(S)=\sum_{v\in S}\phi(v)$.
The goal is to find a minimum-cost edge set such that each component $C$ has $\phi(C)\geq0$; isolated vertices are also components.
The model does not prescribe which supply serves each demand, and supply may remain unused.
The associated set family is $\F=\{S\subseteq V:\phi(S)<0\}$, which is disjointness-compliable because the charges of disjoint sets add.
The GP2P contains the nonfixed point-to-point connection problem, the SF, and rooted $k$-MST as special cases~\citep{journal:even:TALG:05,conference:friggstad:WADS:25}.
For example, for $k\geq1$, assigning charge $-(k-1)$ to a root and charge $1$ to every other vertex requires the root component to contain at least $k$ vertices, including the root.

\citet{journal:goemans:SICOMP:95} obtained a $2$-approximation for the nonfixed P2P.
\citet{journal:even:TALG:05} obtained a logarithmic approximation for a network design problem that includes the GP2P.
\citet{journal:hajiaghayi:TOCS:16} obtained a $2$-approximation when $\phi(V)=0$ and an $\O(\log(\min\{\size{V},\phi(V)+2\}))$-approximation in general; the running time of the latter is pseudo-polynomial because it depends polynomially on the charge values.
\citet{conference:friggstad:WADS:25} removed this pseudo-polynomial dependence and obtained a polynomial-time logarithmic approximation, as well as a $(2+\varepsilon)$-approximation when the number of negative-charge vertices is fixed.

\paragraph{Multiroot Quota Tree.}
In the multiroot quota tree problem (MRQT), each vertex $v$ has a nonnegative prize $b(v)$, and each root $r$ in a set $R\subseteq V$ has a nonnegative quota $k_r$~\citep{preprint:nutov:arXiv:25}.
Each root must belong to a component whose total prize is at least its quota.
Roots in the same component share its prize, so the component needs to meet their largest quota.
For $S\subseteq V$, let $b(S)=\sum_{v\in S}b(v)$ and $k(S)=\max(\{k_r:r\in S\cap R\}\cup\{0\})$.
The goal is to find a minimum-cost edge set whose components satisfy $b(C)\geq k(C)$.
The associated family $\F=\{S\subseteq V:b(S)<k(S)\}$ is disjointness-compliable~\citep{preprint:nutov:arXiv:25}.
The multiroot $k$-MST is the special case with $b(v)=1$ for every vertex and $k_r=k$ for every root.

When $R=V$, every component must contain at least $k$ vertices.
\citet{journal:couetoux:MP:15} gave a $3/2$-approximation for this special case.

\paragraph{Group Steiner Tree.}
In the group Steiner tree problem, we are given a root and a collection $\mathcal{X}$ of vertex groups, and seek a minimum-cost tree containing the root and at least one vertex from each group.
The requirements can be represented by the sets that contain the root but do not intersect some group; this family is disjointness-compliable.

Let $h=\max_{X\in\mathcal{X}}\size{X}$ be the largest group size.
For tree instances, \citet{journal:garg:JALG:00} obtained a randomized $\O(\log\size{\mathcal{X}}\log h)$-approximation.
\citet{conference:halperin:STOC:03} showed that, for every fixed $\varepsilon>0$, even on trees there is no polynomial-time $\log^{2-\varepsilon}\size{\mathcal{X}}$-approximation unless $\mathrm{NP}$ admits quasi-polynomial-time Las Vegas algorithms.
For general graphs, combining the tree algorithm with randomized tree embeddings adds a factor of $\O(\log\size{V})$, resulting in an $\O(\log\size{V}\log\size{\mathcal{X}}\log h)$-approximation~\citep{journal:garg:JALG:00,journal:fakcharoenphol:JCSS:04}.

\paragraph{Covering Steiner.}
The rooted covering Steiner problem extends group Steiner tree by requiring the root component to contain a specified number of vertices from each group.
\citet{journal:gupta:TOC:06} obtained a randomized $\O(\log\size{\mathcal{X}}\log h)$-approximation on trees and a randomized $\O(\rho)$-approximation on general graphs, both with high probability, where $\rho=\O(\log\size{V}\log\size{\mathcal{X}}\log h)$, $\mathcal{X}$ is the collection of groups, and $h=\max_{X\in\mathcal{X}}\size{X}$.
In the multiroot version, the groups in $\mathcal{X}$ may be associated with different roots, and each root's component must satisfy its own group requirements.

\paragraph{DCFC.}
\citet{preprint:nutov:arXiv:25} developed an approximation framework for the DCFC.

Their algorithm contracts the connected components of selected edges and considers a restricted local covering problem for each inclusion-minimal uncovered set.
If membership in every resulting family can be tested in polynomial time and each feasible local covering problem admits an $\alpha$-approximation, their framework achieves ratio $\alpha+\max\{\alpha,2\}\ln\tau$, where $\tau\leq\size{V}$ is the number of inclusion-minimal sets in the initial family.
Their results yield a $3(\ln\size{V'}+1)$-approximation for the GP2P, where $V'$ denotes the set of negative-charge vertices.
For the MRQT, they obtained a $3(\ln\size{R}+1)$-approximation in polynomial time and a $4$-approximation in time exponential in $\size{R}$.
They also obtained a randomized $\O(\rho\log\size{R})$-approximation for multiroot covering Steiner.

\subsection{Our Results}\label{sec:results}
Unless stated otherwise, all approximation algorithms in this paper run in polynomial time.
We first establish a local-to-global approximation framework for the DCFC.
Our local problem is related to the local covering problem of \citet{preprint:nutov:arXiv:25}: it seeks a minimum-cost tree that connects one deficient class to feasible classes so that their union is feasible.
Suppose membership in the family can be decided in polynomial time and this local problem admits an $\alpha$-approximation.
Then, for every fixed $\varepsilon>0$, the DCFC admits a $(1+\varepsilon)(\sqrt{2}+\sqrt{\alpha})^2$-approximation (Theorem~\ref{thm:framework}).

We then apply the framework to the GP2P and the MRQT.
For both problems, the local problem reduces to a rooted quota-tree problem, which admits a $(2+\eta)$-approximation for every fixed $\eta>0$~\citep{conference:friggstad:WADS:25}.
Consequently, for every fixed $\varepsilon>0$, we obtain $(8+\varepsilon)$-approximation algorithms for both problems (Theorems~\ref{thm:gp2p} and~\ref{thm:mrqt}).
To the best of our knowledge, these are the first polynomial-time constant-factor approximations for the GP2P and the MRQT.

We also apply the framework to multiroot covering Steiner and obtain a randomized $\O(\rho)$-approximation (Theorem~\ref{thm:covering-steiner}), improving the previous $\O(\rho\log\size{R})$ ratio of \citet{preprint:nutov:arXiv:25}.

\subsection{Overview of Our Approach}\label{sec:overview}
Our algorithm maintains a partition of the vertices into classes whose vertices have already been connected by selected edges.
A class is deficient if its vertex set belongs to $\F$ and feasible otherwise.
The paths connecting the vertices of a class may pass through other classes.
Thus, a class need not be a connected component of the selected edges.
The local problem considers a region with one deficient class and otherwise feasible classes.
It asks for a minimum-cost tree that connects the deficient class to other classes so that their union is feasible.
The local tree may select only classes in the region, but the paths used to connect them may pass through classes outside the region.

For the GP2P, the deficient class has negative charge, while every other class in the region has nonnegative charge.
We use these nonnegative charges as prizes in a rooted quota-tree instance.
For the MRQT, the deficient class needs additional prize equal to its largest quota minus its current prize.
Each feasible class can contribute its full prize because roots in a component share the prize.
Both local problems can therefore be formulated as rooted quota-tree problems.

To combine the local solutions, we need to control their total cost even when they use overlapping regions.
For this purpose, we consider a nonnegative vector $x$ that assigns a value to each edge.
The value of a cut is the sum of the $x$-values of its edges; these cut values guide the selection of classes and regions.
We first use the cut-relaxation guarantee for SF to connect pairs of deficient vertices whose minimum $x$-cut value exceeds a threshold.
While several deficient classes remain, we use a Gomory--Hu tree~\citep{journal:gomory:JSIAM:61} to choose a region with exactly one deficient class and boundary $x$-value at most the threshold. For the last deficient class, we take the whole graph as the region.
If the union of the classes in the region is
deficient, its boundary yields a violated
cut constraint. Otherwise, the local algorithm
merges the deficient class with selected
feasible classes to form a feasible class.

To bound the total local cost, we compare the construction with a feasible forest.
If no edge of this forest crosses a chosen region, the local tree costs at most $\alpha$ times the cost of the remaining comparison edges inside the region.
After the local connection, all classes in the region are feasible, so we can delete those comparison edges.
Each comparison edge is used in at most one such bound.
The resulting linear inequality holds for the indicator vector of every feasible edge set.
If $x$ satisfies it, we use it to bound the total local cost.
Otherwise, we pass the inequality to the ellipsoid method~\citep{book:groetschel:BOOK:88}.

The remainder of this paper is organized as follows.
Section~\ref{sec:framework} defines the local problem and states the framework theorem. Section~\ref{sec:analysis} presents the construction and analysis. Section~\ref{sec:applications} applies the framework to the GP2P, the MRQT, and multiroot covering Steiner. 

\section{Preliminaries}\label{sec:framework}
Recall that $G=(V,E)$ is an undirected graph with nonnegative edge costs $c(e)$ and $\F$ is the family of sets to be covered.
We assume that all numerical input data are rational for the analysis of the ellipsoid method.
For any $S\subseteq V$, let $\delta(S)$ denote the set of edges with exactly one endpoint in $S$.
An edge covers $S$ if it belongs to $\delta(S)$.
For $F\subseteq E$, let $c(F)=\sum_{e\in F}c(e)$ and let $\chi_F$ be its indicator vector.
For an edge vector $z$ and a set $F\subseteq E$, let $z(F)=\sum_{e\in F}z_e$ and $c^\top z=\sum_{e\in E}c(e)z_e$.
We use $\OPT$ to denote the optimum cost and allow a tree to consist of a single vertex.

\subsection{Set Families and Feasible Components}\label{sec:components}
We first explain why covering $\F$ is equivalent to requiring every component to be feasible.
Recall that $\emptyset\notin\F$ and membership in $\F$ can be decided in polynomial time.
An edge set $F$ covers $\F$ if $F\cap\delta(S)\neq\emptyset$ for every $S\in\F$.
Following \citet{journal:goemans:SICOMP:95,preprint:nutov:arXiv:25}, we call $\F$ disjointness-compliable if
\begin{equation}\label{eq:dc}
S\in\F,\quad B\subseteq S
\quad\Longrightarrow\quad B\in\F\ \text{or}\ S\setminus B\in\F.
\end{equation}
Let $\A=2^V\setminus\F$.
We call a set feasible if it belongs to $\A$, and deficient otherwise.

\begin{lemma}\label{lem:component-equivalence}
Condition~\eqref{eq:dc} holds if and only if the union of two disjoint feasible sets is feasible. In this case, an edge set $F$ covers $\F$ if and only if the vertex set of every connected component of $(V,F)$ is feasible.
\end{lemma}
\begin{proof}
Suppose \eqref{eq:dc} holds, and let $A,B\in\A$ be disjoint.
If $A\cup B\in\F$, then \eqref{eq:dc} implies $A\in\F$ or $B\in\F$, a contradiction.
Hence
\begin{equation}\label{eq:closure}
A,B\in\A,\quad A\cap B=\emptyset
\quad\Longrightarrow\quad A\cup B\in\A.
\end{equation}
Conversely, suppose \eqref{eq:closure} holds.
For $S\in\F$ and $B\subseteq S$, at least one of $B$ and $S\setminus B$ must belong to $\F$.
Otherwise, both sets belong to $\A$, so \eqref{eq:closure} implies $S\in\A$, a contradiction.

For the component condition, we follow the proof of Theorem~2.3 of \citet{journal:goemans:SICOMP:95}. Consider an edge set $F$.
If a component of $(V,F)$ is deficient, no edge of $F$ covers its vertex set, so $F$ does not cover $\F$.
Conversely, suppose every component is feasible but $F$ does not cover a set $S\in\F$.
Since no edge of $F$ crosses $S$, the set $S$ is a union of disjoint feasible components.
By \eqref{eq:closure}, $S$ is feasible, a contradiction.
\end{proof}

By Lemma~\ref{lem:component-equivalence}, the instance is feasible if and only if every connected component $W$ of $G$ belongs to $\A$.
We can therefore handle these components separately and assume that $G$ is connected and $V\in\A$.
Moreover, taking minimum path costs and contracting zero-cost edges preserve $\OPT$, since original solutions remain available at no greater cost, while expanding metric edges or adding zero-cost edges can only merge feasible components, preserving feasibility by \eqref{eq:closure}.
If the zero-cost edges already form a feasible solution, they are optimal.
Hence, we also assume that $G$ is complete, with positive edge costs satisfying the triangle inequality, and that $\OPT>0$.

A natural cut relaxation for the DCFC is
\begin{equation}\label{eq:cut-relaxation}
\min\left\{c^\top x\;\middle|\;x(\delta(S))\geq1\text{ for all }S\in\F,\quad x\geq0\right\}.
\end{equation}
The indicator vector of an edge set satisfies these constraints if and only if the edge set covers $\F$.

\subsection{The Local Problem}\label{sec:local}
After the initial connections, our algorithm handles one deficient class at a time.
Let $\P$ be a partition of $V$ into nonempty sets, called classes.
In our construction, the vertices of each class are connected by already selected edges, possibly through other classes.
For a collection $\S\subseteq\P$, let $U(\S)=\bigcup_{B\in\S}B$ denote the union of its classes.
A class is feasible if it belongs to $\A$ and deficient otherwise.
A collection of classes is feasible if its union is feasible.

Let $c_\P(B,B')$ be the minimum cost of a path between classes $B$ and $B'$ when travel inside each class is free.
For a tree $T$ on classes, let $c_\P(T)=\sum_{BB'\in E(T)}c_\P(B,B')$.
For a region $\S\subseteq\P$, we restrict these costs to pairs of classes in $\S$.
The paths are computed using all classes, so a path representing a tree edge may pass through classes outside the tree, including classes outside $\S$. Note that passing through a class does not by itself select that class.

For a region $\S\subseteq\P$ with exactly one deficient class $A$, the local problem asks for a tree $T$ minimizing $c_\P(T)$ subject to $A\in V(T)\subseteq\S$ and $U(V(T))\in\A$.

Such a tree exists if and only if $U(\S)\in\A$.
In particular, if $U(\S)$ is feasible, a spanning tree of $\S$ is feasible.
Conversely, if a local tree is feasible, its union and all unused classes are feasible, so \eqref{eq:closure} implies that $U(\S)$ is feasible.

\subsection{The Framework Theorem}\label{sec:framework-theorem}
We use the classical $2$-approximation algorithm for the SF~\citep{journal:goemans:SICOMP:95} to connect deficient classes that are difficult to separate by cuts of small value.
For prescribed terminal pairs, a vector $y\geq0$ is feasible for the SF cut relaxation if $y(\delta(S))\geq1$ whenever $S$ separates a terminal pair.
For every such vector $y$, this algorithm returns a forest of cost at most $2c^\top y$.

\begin{theorem}\label{thm:framework}
Suppose that $\F$ is disjointness-compliable and membership in $\F$ can be decided in polynomial time. If the local problem admits an $\alpha$-approximation, then the DCFC admits a $(1+\varepsilon)(\sqrt{2}+\sqrt{\alpha})^2$-approximation for every fixed $\varepsilon>0$.
\end{theorem}

\section{The Local-to-Global Framework}\label{sec:analysis}
We now prove Theorem~\ref{thm:framework}.
For a nonnegative edge vector $x$, our construction first connects pairs of deficient vertices that cannot be separated by cuts of small $x$-value, where a vertex $v$ is deficient if $\{v\}\in\F$.
It then handles regions containing one deficient class at a time.
The construction either returns a feasible forest or an inequality violated by $x$ and satisfied by the indicator vector of every feasible edge set.
We bound the cost of the returned forest in terms of $c^\top x$ and use the ellipsoid method and a budget search to prove the theorem.
Algorithm~\ref{alg:query} summarizes the construction.

We first explain how the algorithm maintains classes and computes the costs of connecting them.

\begin{algorithm}[!t]
\small
\caption{Construction for a given edge vector}
\label{alg:query}
\textbf{Input:} An instance on a metric graph, $x\geq0$, and $0<\theta<1$.\\
\textbf{Output:} A valid inequality violated by $x$, or a feasible forest.
\begin{algorithmic}[1]
\State Group deficient vertices: two belong to the same group if their minimum $x$-cut value exceeds $\theta$.
\State Use one SF call to connect the vertices within every group; let $H$ be the returned edge set.
\State Let $\P$ be the partition formed by the groups and the remaining singleton vertices.
\State Set $i\coloneq0$.
\While{a deficient class remains}
\State Set $i\coloneq i+1$.
\State If only one class is deficient, take $\S_i=\P$. Otherwise, choose $\S_i$ by the Gomory--Hu procedure below.
\State Let $S_i\coloneq U(\S_i)$ and let $A_i$ be the unique deficient class in $\S_i$.
\If{$S_i$ is deficient}
\State \Return $z(\delta(S_i))\geq1$.
\EndIf
\State Apply the local algorithm to $\S_i$ and $A_i$; let $T_i$ be the returned tree and $a_i=c_\P(T_i)$.
\State Add paths representing the edges of $T_i$ to $H$ and update $\P$ by merging the classes in $V(T_i)$.
\EndWhile
\State Set $t\coloneq i$.
\If{$\sum_{j=1}^{t}a_j(1-x(\delta(S_j)))>\alpha c^\top x$}
\State \Return inequality~\eqref{eq:deletion}, with the sets $S_j$ and costs $a_j$ fixed.
\EndIf
\State Remove cycles from $H$ and \Return $H$.
\end{algorithmic}
\end{algorithm}

\subsection{Classes}\label{sec:logical}
Let $H$ be the set of selected edges and let $\P$ be the current partition into classes.
The vertices of each class are connected by edges in $H$, but their connecting paths may pass through other classes.
We merge only the classes explicitly selected by the algorithm.
Thus, a class need not be a connected component of $(V,H)$, and a path may connect additional classes without merging them.

Recall that $c_\P$ is the cost between classes when travel inside each class is free.
For an edge $uv\in E$ whose endpoints lie in distinct classes $B$ and $B'$, we have $c_\P(B,B')\leq c(uv)$.
To connect two classes at cost $c_\P(B,B')$, we add the edges of a corresponding path and use paths already in $H$ to connect vertices belonging to the same class.
The cost of the added edges is at most $c_\P(B,B')$.
Each component of $(V,H)$ nevertheless contains whole classes.
Thus, when every class is feasible, every component of $(V,H)$ is feasible by \eqref{eq:closure}.

We use the values of $x$ to choose the classes to connect.
Fix $x\geq0$ on $E$ throughout one construction.
For two distinct classes, we sum the $x$-values of the edges joining them; edges within one class do not affect these cut values.
Thus, for every $\S\subseteq\P$, the cut separating $\S$ from the other classes has value $x(\delta(U(\S)))$.
These cut values are separate from the costs $c_\P$ used by the local algorithm.

\subsection{Initial Connections}\label{sec:initial}
Fix a threshold $0<\theta<1$ and let $P=c^\top x$.
We first connect deficient vertices that cannot be separated by a cut of $x$-value at most $\theta$.
This step will allow us to find a region with just one deficient class.
Two deficient vertices belong to the same group if their minimum $x$-cut value exceeds $\theta$.
Equivalently, they lie on the same side of every cut of $x$-value at most $\theta$; hence, this relation partitions the deficient vertices into groups.

For each group, we choose a representative and require it to be connected to every other vertex of the group.
Every cut separating one of these pairs has $x$-value greater than $\theta$.
Thus, $x/\theta$ is feasible for the SF cut relaxation for all these pairs, and the SF algorithm finds an edge set $H_0$ with
\begin{equation}\label{eq:initial-cost}
c(H_0)\leq\frac{2}{\theta}P.
\end{equation}
We set $H=H_0$ and merge only the deficient vertices within each group.
All other vertices remain separate classes, even if they are connected by edges in $H_0$.
A merged group may already be feasible; we only need to handle the classes that remain deficient.

For any two distinct groups, some cut of $x$-value at most $\theta$ separates a vertex of one from a vertex of the other.
Such a cut cannot split any group, so it remains a cut of the same value after the merges.
Hence, every pair of remaining deficient classes can be separated by a cut of $x$-value at most $\theta$.

\subsection{A Region with One Deficient Class}\label{sec:regions}
We next select a region for the local problem.
We need its boundary to have a small $x$-value: if the region is deficient, its boundary will yield a violated cut constraint.
We also need the remaining deficient classes to stay separable by cuts of $x$-value at most $\theta$ after a local connection.

\begin{lemma}\label{lem:region}
Suppose at least one class is deficient and each pair of deficient classes has a separating cut of $x$-value at most $\theta$.
We can find in polynomial time a region with exactly one deficient class and boundary $x$-value at most $\theta$.
After any merge within this region, every pair of deficient classes outside it can still be separated by a cut of $x$-value at most $\theta$.
\end{lemma}
\begin{proof}
If only one deficient class remains, we use all current classes as the region.
Its boundary is empty, and there are no outside deficient classes to consider.
Otherwise, we construct a Gomory--Hu tree for the current classes~\citep{journal:gomory:JSIAM:61}.
The capacity between two classes is the sum of the $x$-values of the edges joining them.
The cut induced by each tree edge has $x$-value equal to the weight of that edge, and the minimum edge weight on the tree path between two classes is their minimum-cut value.
We contract all tree edges of weight greater than $\theta$ only for the purpose of selecting the region.
Every resulting tree vertex contains at most one deficient class; otherwise, the path between two such classes would have minimum weight greater than $\theta$.

We mark the tree vertices containing deficient classes and take the minimal subtree spanning them.
Each leaf of this subtree is marked.
We choose a leaf and remove its unique edge toward the rest of the subtree.
Let $\S$ be the classes on the side containing that leaf, including the classes contracted when selecting the region.
This side contains exactly one deficient class.
The removed edge has weight at most $\theta$, so $x(\delta(U(\S)))\leq\theta$.
An illustration of this choice of the region is shown in Figure~\ref{fig:region-selection}.

\begin{figure}[t]
\centering
\begin{tikzpicture}[x=1.2cm,y=0.85cm,
  marked/.style={circle,draw,fill=black,minimum size=7pt,inner sep=0pt},
  unmarked/.style={circle,draw,fill=white,minimum size=7pt,inner sep=0pt}]
  
  \draw[dashed,rounded corners=4pt] (-3.65,-1.30) rectangle (-1.25,0.68);
  \node at (-2.45,1.08) {$\S$};

  \draw (-3.1,0) -- (-1.8,0) -- (0,0)
        (-1.8,0) -- (-2.0,-0.9)
        (0,0) -- (0.95,1.1) -- (2.1,1.1)
        (0,0) -- (0.95,-1.1) -- (2.1,-1.1);
  \draw[very thick] (-1.8,0) -- (0,0);

  \node[unmarked] at (-3.1,0) {};
  \node[marked] at (-1.8,0) {};
  \node[unmarked] at (-2.0,-0.9) {};
  \node[unmarked] at (0,0) {};
  \node[unmarked] at (0.95,1.1) {};
  \node[marked] at (2.1,1.1) {};
  \node[unmarked] at (0.95,-1.1) {};
  \node[marked] at (2.1,-1.1) {};

  \node[above] at (-1.8,0.08) {$A$};
  \node[above] at (2.1,1.18) {$B$};
  \node[below] at (2.1,-1.18) {$C$};
  \node[above] at (-0.85,0.04) {$e$};
  \node[below] at (-0.85,-0.04) {$\leq\theta$};
\end{tikzpicture}
\caption{Black vertices contain deficient classes in the contracted Gomory--Hu tree, and removing edge $e$ separates $\S$ from the other deficient classes.}
\label{fig:region-selection}
\end{figure}
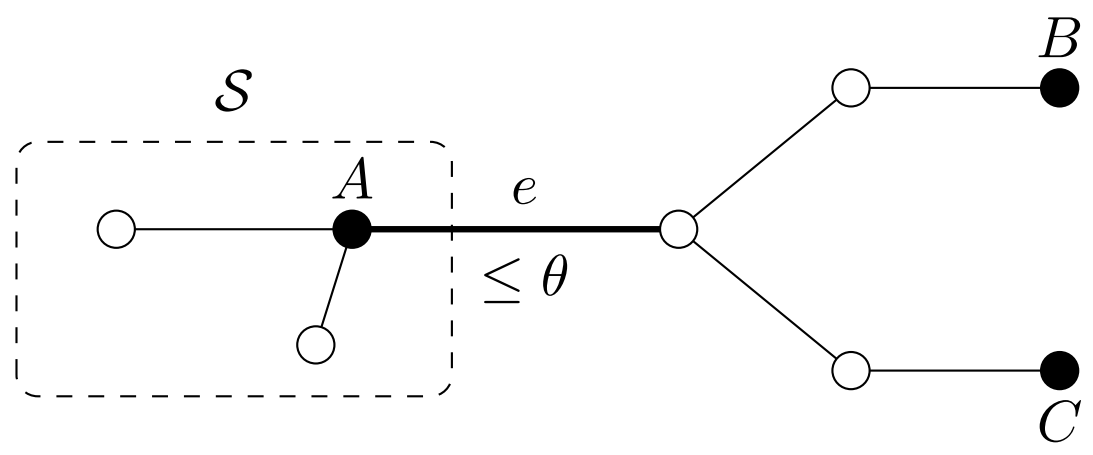

The tree path between any two deficient classes outside this side avoids it.
That path contains an edge of weight at most $\theta$ by the assumed pairwise property.
The cut represented by this edge places all of $\S$ on one side.
Any merge within $\S$ leaves this cut unchanged.
It still separates the two outside deficient classes and has $x$-value at most $\theta$.
\end{proof}

At step $i$, we select a region $\S_i$ using Lemma~\ref{lem:region}.
Let $S_i=U(\S_i)$ be the set of vertices in these classes and let $A_i$ be their unique deficient class.
If $S_i\notin\A$, we return the inequality $z(\delta(S_i))\geq1$.
By Lemma~\ref{lem:component-equivalence}, this inequality holds for every feasible edge set, but it is violated by $x$ because $x(\delta(S_i))\leq\theta<1$.

If $S_i\in\A$, the local problem is feasible.
We apply the local $\alpha$-approximation to $\S_i$ with root class $A_i$.
Let $T_i$ be the returned tree and $a_i$ its cost under $c_\P$.
We add paths representing its edges to $H$ and merge exactly the classes in $V(T_i)$.
Their union is feasible by the local guarantee.
Every unselected class in $\S_i$ was already feasible, so every class represented in $S_i$ is now feasible.
The number of deficient classes decreases by one, and Lemma~\ref{lem:region} permits the next step.

Unless a violated cut constraint is returned, the construction ends after $t\leq\size{V}$ steps.
Every final class is feasible, so the selected edge set $H$ is feasible.
Its cost satisfies
\begin{equation}\label{eq:purchase-cost}
c(H)\leq c(H_0)+\sum_{i=1}^{t}a_i
\leq\frac{2}{\theta}P+\sum_{i=1}^{t}a_i.
\end{equation}

\subsection{Bounding the Cost of the Local Trees}\label{sec:deletion}
Fix a completed construction, including its regions, trees, and merges.
To bound the total cost of the local trees, we compare the construction with a feasible edge set.
For an edge vector $z$, consider the following inequality:
\begin{equation}\label{eq:deletion}
\sum_{i=1}^{t}a_i\bigl(1-z(\delta(S_i))\bigr)\leq\alpha c^\top z.
\end{equation}
For each region with no comparison edge crossing its boundary, we use the remaining comparison edges inside the region to bound the cost of its local tree.
We then delete these edges so that they cannot be used again.

\begin{lemma}\label{lem:deletion}
For every completed construction, inequality~\eqref{eq:deletion} holds at $z=\chi_F$ for every feasible edge set $F$.
\end{lemma}
\begin{proof}
Fix a feasible edge set $F$ and let $I=\{i:F\cap\delta(S_i)=\emptyset\}$.
The definition of $I$ uses all edges of $F$, including those deleted in earlier steps.
Let $L_i\subseteq F$ be the remaining comparison edges just before step $i$, with $L_1=F$.
We form a graph whose vertices are the current classes, with an edge for each edge of $L_i$ joining the classes containing its endpoints.
We maintain the invariant that the vertices represented by each component of the resulting graph form a feasible set.
Initially, each such component is a union of components of $(V,F)$, so it is feasible by \eqref{eq:closure}.
Edges within a class become loops and do not affect connectivity.

First consider $i\in I$.
Since $L_i\subseteq F$, no remaining comparison edge crosses $S_i$.
The component containing $A_i$ in this graph therefore lies entirely in $\S_i$.
The union of its classes is feasible by the invariant, so a spanning tree of this component is a feasible solution to the local problem.
Let $D_i$ be all edges of $L_i$ with both endpoints in $S_i$.
By the definition of $c_\P$, the component has a spanning tree of cost at most $c(D_i)$.
The local optimum is at most $c(D_i)$, and hence
\begin{equation}\label{eq:local-comparison}
a_i\leq\alpha c(D_i).
\end{equation}
An illustration of this step is shown in Figure~\ref{fig:comparison-deletion}.

\begin{figure}[t]
\centering
\begin{tikzpicture}[
  x=0.95cm,y=0.76cm,
  every node/.style={font=\small},
  class/.style={circle,draw,minimum size=7mm,inner sep=0pt,fill=white},
  merged/.style={draw,rounded corners=8pt,minimum height=7mm,inner sep=2pt,fill=white}
]

\begin{scope}
  \draw[densely dotted,rounded corners=4pt] (0,0) rectangle (4.35,2.45);
  \node[anchor=north west] at (0.06,2.40) {$S_i$};
  \node[class,fill=black!18] (a) at (0.80,1.45) {$A_i$};
  \node[class] (b) at (3.55,1.45) {$B$};
  \node[class] (c) at (2.18,0.60) {$C$};
  \draw[thick] (a) -- (c) -- (b);
  \draw[dashed] (a) -- node[above] {$T_i$} (b);
  \node[class] (p) at (5.12,1.45) {$P$};
  \node[class] (q) at (6.15,1.45) {$Q$};
  \draw[thick] (p) -- (q);
  \node[anchor=north] at (3.10,-0.15) {(a) Before the merge};
\end{scope}

\begin{scope}[xshift=7.1cm]
  \draw[densely dotted,rounded corners=4pt] (0,0) rectangle (4.35,2.45);
  \node[anchor=north west] at (0.06,2.40) {$S_i$};
  \node[merged] (ab) at (1.12,1.45) {$A_i\cup B$};
  \node[class] (c2) at (3.23,1.45) {$C$};
  \node[class] (p2) at (5.12,1.45) {$P$};
  \node[class] (q2) at (6.15,1.45) {$Q$};
  \draw[thick] (p2) -- (q2);
  \node[anchor=north] at (3.10,-0.15) {(b) After deleting $D_i$};
\end{scope}
\end{tikzpicture}
\caption{When $F\cap\delta(S_i)=\emptyset$, merging the classes selected by the dashed local tree $T_i$ makes all classes in $S_i$ feasible, allowing the solid comparison edges inside $S_i$ to be deleted.}
\label{fig:comparison-deletion}
\end{figure}
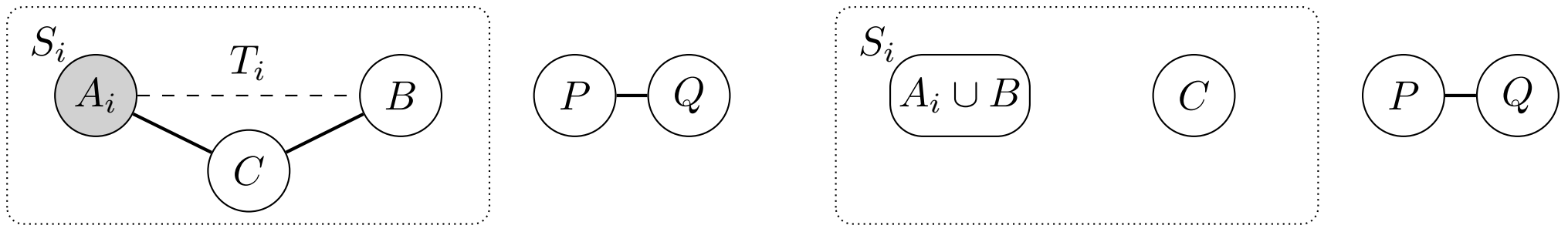

After the merge, set $L_{i+1}=L_i\setminus D_i$ if $i\in I$ and $L_{i+1}=L_i$ otherwise.
For $i\in I$, no comparison edge crosses $S_i$, so deleting $D_i$ leaves the outside components unchanged.
Each resulting class in $S_i$ is feasible and isolated.
For $i\notin I$, no edge is deleted, and the vertices represented by each new component form the union of those represented by whole previous components.
This union is feasible by \eqref{eq:closure}.
Thus, the invariant is preserved.

An edge is removed as soon as it belongs to a set $D_i$ with $i\in I$.
These sets are pairwise disjoint even when their regions overlap.
Equation~\eqref{eq:local-comparison} therefore implies
\begin{equation}\label{eq:disjoint-payments}
\sum_{i\in I}a_i
\leq\alpha\sum_{i\in I}c(D_i)
\leq\alpha c(F).
\end{equation}
For every $i\notin I$, $\size{F\cap\delta(S_i)}\geq1$, so its term in the left-hand side of \eqref{eq:deletion} is nonpositive at $z=\chi_F$.
Equation~\eqref{eq:disjoint-payments} then proves \eqref{eq:deletion}.
\end{proof}

The edge set $F$ and the deleted sets $D_i$ are used only in the proof; the algorithm does not need to find them.
In particular, $F$ need not contain any edge selected by the algorithm.

\subsection{Separation or a Feasible Forest}\label{sec:query}
After completing the construction, we test \eqref{eq:deletion} at $z=x$.
If the inequality is violated, we return it with the sets $S_i$ and costs $a_i$ fixed.
Otherwise, every selected region has boundary $x$-value at most $\theta$, and
\begin{equation}\label{eq:local-total}
(1-\theta)\sum_{i=1}^{t}a_i
\leq\sum_{i=1}^{t}a_i\bigl(1-x(\delta(S_i))\bigr)
\leq\alpha P.
\end{equation}
Equations~\eqref{eq:purchase-cost} and \eqref{eq:local-total} imply
\begin{equation}\label{eq:query-factor}
c(H)\leq\left(\frac{2}{\theta}+\frac{\alpha}{1-\theta}\right)c^\top x.
\end{equation}
We remove cycles from $H$ to obtain a forest with the same components and no greater cost.

Every cut constraint returned by the construction has nonnegative coefficients on the left-hand side.
We can also express \eqref{eq:deletion} in this form:
\begin{equation}\label{eq:nonnegative-row}
\sum_{e\in E}\left(\alpha c(e)+\sum_{i:e\in\delta(S_i)}a_i\right)z_e
\geq\sum_{i=1}^{t}a_i.
\end{equation}
We use this property to show that the ellipsoid method finds a feasible forest when the budget is large enough.

Each application of the construction uses one SF call, at most $\size{V}$ local calls, and polynomially many minimum-cut and shortest-path computations.
The membership tests and all tree subroutines are polynomial-time by assumption, and each returned inequality has a polynomial-size description.
Thus, Algorithm~\ref{alg:query} runs in polynomial time.

\subsection{Finding a Suitable Vector}\label{sec:ellipsoid}
We now use the ellipsoid method to find a vector for which Algorithm~\ref{alg:query} returns a feasible forest.
We do not require an exact separation algorithm for the full cut relaxation.
For a budget $U>0$, we apply the ellipsoid method to $z\geq0$ and $c^\top z\leq U$~\citep{book:groetschel:BOOK:88}.
If a vector violates these constraints, we return one of them.
For a vector $x$ satisfying these constraints, we run Algorithm~\ref{alg:query}.
Each call starts a new construction; if it returns an inequality, we discard its selected edges.
We continue if the algorithm returns a violated inequality and stop if it returns a feasible forest.

\begin{lemma}\label{lem:budget}
For every fixed $0<\zeta<1$, the procedure for a budget $U$ runs in polynomial time.
If $U\geq(1+\zeta)\OPT$, it returns a forest of cost at most $(2/\theta+\alpha/(1-\theta))U$.
\end{lemma}
\begin{proof}
Let $m=\size{E}$, let $c_{\min}=\min_{e\in E}c(e)>0$, and set $R_0=U/c_{\min}$.
These constraints imply $z\in[0,R_0]^E$.
We start with an ellipsoid containing this box.
In the procedure for budget $U$, every vector for which the algorithm returns a forest satisfies $c^\top x\leq U$, so \eqref{eq:query-factor} proves the claimed cost bound.

Suppose $U\geq(1+\zeta)\OPT$ and fix an optimal edge set $F^*$.
To show that the method must find a forest, we identify a box that satisfies the budget constraints and all inequalities returned by Algorithm~\ref{alg:query}.
Set $\mu=\zeta U/(8\sum_{e\in E}c(e))$ and consider the box $\B=\chi_{F^*}+[\mu,2\mu]^E$.
Every $z\in\B$ is nonnegative and satisfies
\begin{equation}\label{eq:protected-budget}
c^\top z\leq\OPT+2\mu\sum_{e\in E}c(e)
=\OPT+\frac{\zeta U}{4}
\leq\frac{U}{1+\zeta}+\frac{\zeta U}{4}
\leq U.
\end{equation}
Every inequality returned by Algorithm~\ref{alg:query} is satisfied by $\chi_{F^*}$ and can be written with nonnegative coefficients on the left-hand side.
It is therefore satisfied by every vector whose coordinates are at least those of $\chi_{F^*}$, including every vector in $\B$.
Thus, every inequality returned to the ellipsoid method preserves the same box of volume $\mu^m$.

The standard ellipsoid volume bound reduces the enclosing volume below $\mu^m$ after $\O(m^2\log(2mR_0/\mu))$ cuts.
This cannot happen while $\B$ is preserved.
We stop after this many iterations and report failure if no forest has been found.
When $U\geq(1+\zeta)\OPT$, failure is therefore impossible.
Since $R_0/\mu=8\sum_{e\in E}c(e)/(\zeta c_{\min})$, the iteration bound is polynomial in the input size for fixed $\zeta$.
Each returned inequality also has a polynomial-size description, so the standard polynomial-time implementation of the ellipsoid method applies.
We do not impose upper bounds $z_e\leq1$, since they would exclude the box $\B$.
\end{proof}

Failure for a budget $U$ does not certify that $\OPT>U$.
Instead, we use the lemma to find a sufficiently large budget.
Let $T_0$ be a spanning tree of $G$.
It is feasible because the input instance is feasible, and $c_{\min}\leq\OPT\leq c(T_0)$.
We test $U_j=c_{\min}(1+\zeta)^j$ in increasing order until success, stopping no later than the first budget at least $(1+\zeta)c(T_0)$.
The first budget at least $(1+\zeta)\OPT$ is at most $(1+\zeta)^2\OPT$ and must succeed by Lemma~\ref{lem:budget}.
The number of tested budgets is polynomial, and the first successful forest satisfies
\begin{equation}\label{eq:outer-factor}
c(H)\leq(1+\zeta)^2\left(\frac{2}{\theta}+\frac{\alpha}{1-\theta}\right)\OPT.
\end{equation}

The factor $2/\theta+\alpha/(1-\theta)$ is minimized at $\theta^*=\sqrt{2}/(\sqrt{2}+\sqrt{\alpha})$, where it equals $(\sqrt{2}+\sqrt{\alpha})^2$.
For fixed $\varepsilon>0$, choose a rational $0<\zeta<1$ with $(1+\zeta)^3<1+\varepsilon$. We choose a rational threshold sufficiently close to $\theta^*$ so that the factor is at most $1+\zeta$ times its minimum.
Equation~\eqref{eq:outer-factor} then yields a ratio strictly below $(1+\varepsilon)(\sqrt{2}+\sqrt{\alpha})^2$. We can therefore use a sufficiently close rational upper bound on $\alpha$ in the cost test and still obtain the stated ratio. This proves Theorem~\ref{thm:framework}.

\section{Applications}\label{sec:applications}
We now apply the framework to the GP2P, the MRQT, and multiroot covering Steiner.
For the first two problems, the local problem reduces to a rooted quota-tree problem, which admits a $(2+\eta)$-approximation for every fixed $\eta>0$~\citep{conference:friggstad:WADS:25}.
For each local region $\S$, the quota-tree instance has vertex set $\S$ and edge costs $c_\P$.
Rational prizes and quotas can be scaled to integers without changing the feasible trees or their costs.

\subsection{Generalized Point-to-Point Connection}\label{sec:gp2p}
Recall that in the GP2P, each vertex $v$ has an integer charge $\phi(v)$, and every component $C$ must satisfy $\phi(C)\geq0$.
Each contracted vertex has the total charge of the original vertices it represents.
Let $\A=\{S\subseteq V:\phi(S)\geq0\}$.
We can check whether a set belongs to $\A$ by summing its charges.
For disjoint $A,B\in\A$, we have $\phi(A\cup B)=\phi(A)+\phi(B)\geq0$; thus, \eqref{eq:closure} holds.
For the local problem, each class retains the total charge of its vertices.

\begin{theorem}\label{thm:gp2p}
For every fixed $\varepsilon>0$, the GP2P admits an $(8+\varepsilon)$-approximation.
\end{theorem}
\begin{proof}
Consider a local region $\S$ with one deficient class $A$.
We have $\phi(A)<0$, while every other class $B\in\S$ has nonnegative charge.
To make the union of $A$ and the selected classes feasible, we need to collect a total charge of at least $-\phi(A)$ from these other classes.
Thus, we construct a rooted quota-tree instance with root $A$, root prize zero, prize $\phi(B)$ at each other class $B$, and quota $-\phi(A)$.
A tree is feasible for this instance if and only if the union of its classes has nonnegative charge.
The two local problems have the same feasible trees and costs.
Therefore, the rooted quota-tree algorithm meets the local requirement with factor $2+\eta$ for every fixed $\eta>0$.
The result follows from Theorem~\ref{thm:framework} by choosing $\eta$ and the theorem's accuracy parameter sufficiently small.
\end{proof}



\subsection{Multiroot Quota Tree}\label{sec:mrqt}
Recall that in the MRQT, each vertex has a nonnegative prize $b(v)$, and each root $r\in R$ has a quota $k_r$.
Every component $C$ must satisfy $b(C)\geq k(C)$, where $b(C)$ is its total prize and $k(C)$ is the largest quota of a root in $C$, or zero if $C$ contains no root.
Each contracted vertex retains its total prize and largest root quota.
Let $\A=\{S\subseteq V:b(S)\geq k(S)\}$.
We can check membership in $\A$ in polynomial time.
For disjoint feasible sets $A$ and $B$, their union has at least as much prize as either set and its largest quota is $\max\{k(A),k(B)\}$.
Thus, $A\cup B$ is feasible and \eqref{eq:closure} holds.
For the local problem, each class retains its total prize and its largest root quota.

\begin{theorem}\label{thm:mrqt}
For every fixed $\varepsilon>0$, the MRQT admits an $(8+\varepsilon)$-approximation.
\end{theorem}
\begin{proof}
Consider a local region $\S$ with one deficient class $A$.
Class $A$ needs at least $k(A)-b(A)>0$ additional prize to meet its largest quota.
We therefore construct a rooted quota-tree instance with root $A$, root prize zero, prize $b(B)$ at each other class $B\in\S$, and quota $k(A)-b(A)$.

Let $C$ be the union of the classes in a tree that meets this quota.
Then $b(C)\geq k(A)$.
Every other selected class $B$ was already feasible, so $b(C)\geq b(B)\geq k(B)$.
Thus, $b(C)\geq k(C)$, even if a selected class has a larger quota than $A$.
Conversely, if the union $C$ of a tree is feasible, then $b(C)\geq k(C)\geq k(A)$, so the tree meets the quota.
Hence, the two local problems have the same feasible trees and costs.
The rooted quota-tree algorithm meets the local requirement with factor $2+\eta$ for every fixed $\eta>0$.
The result follows from Theorem~\ref{thm:framework} by choosing $\eta$ and the theorem's accuracy parameter sufficiently small.
\end{proof}

\begin{corollary}\label{cor:multiroot-kmst}
For every fixed $\varepsilon>0$, multiroot $k$-MST admits an $(8+\varepsilon)$-approximation, with no restriction on the number of roots or on $k$.
\end{corollary}
\begin{proof}
Set $b(v)=1$ for all original vertices and $k_r=k$ for all roots in Theorem~\ref{thm:mrqt}.
The component condition becomes $\size{C}\geq k$ whenever $C\cap R\neq\emptyset$.
\end{proof}

\subsection{Multiroot Covering Steiner}\label{sec:covering-steiner}
Recall that in the multiroot covering Steiner problem, each root $r\in R$ has a collection of groups $\mathcal{X}_r$, and its component must contain at least $k_{r,X}$ vertices from each group $X\in\mathcal{X}_r$.
Let $\mathcal{X}=\bigcup_{r\in R}\mathcal{X}_r$ and $h=\max_{X\in\mathcal{X}}\size{X}$.
A set is feasible if it satisfies the requirements of every root it contains.
Membership can be checked in polynomial time, and the union of disjoint feasible sets is feasible.

\begin{theorem}\label{thm:covering-steiner}
Multiroot covering Steiner admits a randomized $\O(\rho)$-approximation with high probability, where $\rho=\O(\log\size{V}\log\size{\mathcal{X}}\log h)$.
\end{theorem}
\begin{proof}
Consider a local region $\S$ with one deficient class $A$.
Since every other class in $\S$ is feasible, it suffices to satisfy the requirements of the roots in $A$.
For each group required by a root in $A$, we use the largest requirement among the roots in $A$ that require this group.

We construct a rooted covering Steiner instance with one vertex for each class in $\S$, root $A$, and edge costs $c_\P$.
For each original vertex represented by a class in $\S$, we add a leaf joined to that class by a zero-cost edge.
For each required group $X$, we include the leaves corresponding to its original vertices represented in $\S$, with the requirement defined above.
The instance has at most twice as many vertices as the input graph, at most $\size{\mathcal{X}}$ groups, and group sizes at most $h$.

A feasible local tree can be extended at no cost by adding all leaves of its classes, yielding a feasible covering Steiner tree.
Conversely, the classes of a feasible covering Steiner tree form a feasible local tree of no greater cost.
Thus, the two optima are equal, and the randomized covering Steiner algorithm gives a local $\O(\rho)$-approximation with high probability~\citep{journal:gupta:TOC:06}.

The framework makes polynomially many local calls, including those made during the ellipsoid procedure and budget search.
We repeat the local algorithm independently and select the cheapest feasible tree, using a spanning tree of $\S$ if no feasible tree is returned.
With a polynomial number of repetitions, a union bound ensures that all local calls satisfy the $\O(\rho)$ guarantee with high probability.
On this event, the analysis of the framework applies and returns a feasible forest with the claimed ratio.
If the budget search finds no forest, we return a spanning tree of $G$.
\end{proof}

\section*{Acknowledgement}
OpenAI's ChatGPT (GPT-6 Pro) was used for research exploration, proof checking, and manuscript preparation.
The authors independently verified the results, reorganized the paper's structure, rewrote the manuscript, and take full responsibility for its content.

\bibliographystyle{apalike}
\bibliography{L2G}
\end{document}